# Chiral Coherent Perfect Absorption on Exceptional Surfaces

S. Soleymani,[1] Q. Zhong,[2] M. Mokim,[1] S. Rotter,[3] R. El-Ganainy,[2] & Ş. K. Özdemir,[1,4]

[1] *Department of Engineering Science and Mechanics, The Pennsylvania State University, University Park, PA 16802, USA*

[2] *Department of Physics and Henes Center for Quantum Phenomena, Michigan Technological University, Houghton, Michigan, 49931, USA*

[3] *Institute for Theoretical Physics, Vienna University of Technology, A–1040 Vienna, Austria*

[4] *Materials Research Institute (MRI), The Pennsylvania State University, University Park, PA 16802, USA*

**Engineering the transport of radiation and its interaction with matter using non-Hermiticity, particularly through spectral degeneracies known as exceptional points (EPs), is an emerging field that has both fundamental and practical implications. Chiral behaviour in the vicinity of EPs opens new opportunities in radiation control, such as unidirectional reflection or lasing with potential applications in areas ranging from cavity quantum electrodynamics and spectral filtering, to sensing and thermal imaging. However, tuning and stabilizing a system to a discrete EP in parameter space is a challenging task—either the system is operated close to an EP rather than directly at the EP or the true power of EPs is obscured by stability issues. Here, we circumvent this challenge by designing a photonic system that operates on a surface of exceptional points, known as an exceptional surface (ES). We achieve this by using a waveguide-coupled optical resonator with an external feedback loop that induces a nonreciprocal coupling between its frequency degenerate clockwise (CW) and counter-clockwise (CCW) traveling modes. By operating the system at critical coupling on the ES, we demonstrate, for the first time, the effect of chiral and degenerate coherent perfect absorption (CPA) where input waves in only one direction are perfectly absorbed. This novel type of CPA-EP is revealed through the observation of the predicted and long-sought squared Lorentzian absorption lineshape. We expect our results and approach to serve as a bridge between non-Hermitian physics and other fields that rely on radiation engineering.**

Exceptional points (EPs) are generic degeneracies of non-Hermitian systems, where two or more eigenvalues and the associated eigenvectors of a system coalesce, resulting in the reduction of the system's dimensionality and in a severely skewed vector space[1-6]. This is very different from Hermitian degeneracies known as diabolic points (DPs) where eigenvectors stay orthogonal to each other although the eigenvalues are degenerate[1-8]. This difference has created a variety of novel opportunities and attracted enormous attention from different scientific disciplines. Among the intriguing phenomena at or in the vicinity of EPs are chiral behaviour[9], enhanced response to small perturbations[10-15], and enhanced transmission and lasing with increasing loss[16-18], just to name a few [see references 2-6 for a complete list].

EPs emerge in systems through different routes, such as balanced loss and gain as in parity-time (PT) symmetric systems[19-22], judiciously engineered loss imbalance in loss-only systems[16,17,23,24], asymmetric coupling between modes of a system (e.g., CW and CCW modes)[9,10], post-selection in quantum systems[25,26], and parametric modulation[27], all of which have been demonstrated in experiments. The latter three routes differ from the former two routes because they do not rely on introducing additional loss or gain into the system, and thus remain free from the associated noise contributions. As such they have the potential to be used in sensors[11,28,29], spontaneous emission control[30,31], and other fields where noise imposes stringent constraints.

The asymmetric coupling between CW and CCW modes of waveguide-coupled microring resonators has been achieved in experiments through the control of intermodal scattering by inserting two scatterers in the resonator mode field whose size and relative distance in the field can be tuned[9,24]. This procedure gives rise to isolated exceptional points that are typically very difficult to stabilize against fabrication errors and fluctuations in the experimental environment. To overcome this problem, the concept of exceptional surfaces (ES) was recently introduced[28] with potential applications in optical sensing[28], optical amplification[32], and spontaneous emission control[30]: these hypersurfaces in parameter space consist of a continuous collection of exceptional points. An exceptional surface emerges in waveguide-coupled ring resonators through unidirectional (i.e., non-reciprocal) coupling between the modes such that CW mode couples to the CCW but the CCW mode does not couple to CW mode or vice versa (**Fig. 1a, b**). This feature provides a stability against unwanted perturbations (e.g., noise, fabrication imperfections, etc.) that typically drive the system away from discrete EPs and thereby deteriorate their most desirable features.

While, in principle, unidirectional coupling in microring resonators can be achieved via back-reflection from a simple end-mirror placed at one of the output ends of the waveguide[28] (see **Fig. 1a**), we take a different route here that provides full experimental control over the strength and phase of the back-reflected signal (i.e., we can tune the reflection magnitude and phase of the end-mirror or the reflector), and hence allows us to steer the system on the ES. By critical coupling to the resonator on the ES we also observe, for the first time, the novel type of perfectly absorbing EP[33] with the unique features of chiral absorption[34] (i.e., higher absorption for incidence from a specific direction) and a characteristic quartic absorption lineshape[33]. Since this effect of degenerate CPA takes place here on the ES, we refer to it as CPA-ES.

Our experimental system is composed of a whispering gallery mode (WGM) resonator coupled to a tapered-fiber waveguide, which is used to couple light in and out of a resonant mode, and a feedback loop acting as a tunable end-mirror (**Fig. 1a**). The resonator supports CW and CCW modes at the same frequency. We have identified a resonance mode with intrinsic quality factor of $1.4 \times 10^6$ (measured in the regime of deep undercoupling) and confirmed that in the absence of the feedback loop, the transmission ($T_{cw}$ detected at $D_2$ and $T_{ccw}$ detected at $D_1$) and reflection spectra ($R_{cw}$ detected at $D_1$ and $R_{ccw}$ detected at $D_2$) are symmetric for light input in the CW direction (forward or left incidence) and CCW direction (backward or right incidence) (see **Fig. 1c**): both $T_{cw}$ and $T_{ccw}$ exhibit typical Lorentzian lineshapes, and $R_{cw} = R_{ccw} = 0$, implying that there is no intermodal coupling between the CW and CCW modes (i.e., no mode splitting). We have also confirmed that the loading curve is the same for left and right incidence (inputs in the CW and CCW directions) (see **Fig. 1c, inset**). The introduction of the feedback loop acting as a tunable end-mirror breaks this symmetry by inducing unidirectional coupling between the CW and CCW modes: light transmitted in the forward direction (left incidence; CW mode) is reflected back and couples into the CCW mode, but light transmitted in the opposite direction (right incidence; CCW mode) travels directly to a detector without any back-coupling into the CW mode (i.e., no feedback loop or reflector at the output of the taper in the CCW direction).

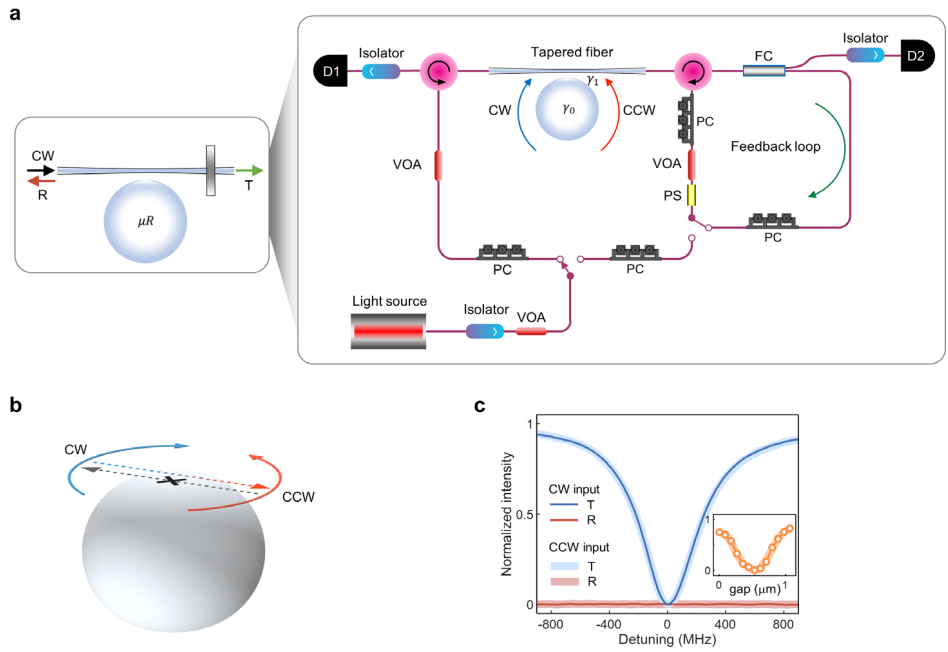

**Figure 1 | Experimental setup for tunable unidirectional coupling between CW and CCW modes of a ring resonator. a**, A partially reflecting mirror placed at only one end of a waveguide (end-mirror) in a waveguide-coupled microresonator ($\mu R$) system unidirectionally couples the frequency-degenerate clockwise (CW) and counter-clockwise (CCW) modes of the resonator. Such an end-mirror with tunable reflection coefficient is implemented using a feedback loop in which the magnitude and phase of the feedback is electrically tuned using a variable optical attenuator (VOA) and phase shifter (PS). A 1-to-2 fiber coupler (FC) placed before the feedback loop provides a tap to monitor a portion of the transmission for an input in the CW direction (i.e., a portion of the $T_{cw}$). $\gamma_1$: waveguide-resonator coupling loss; $\gamma_0$: sum of all resonator related losses such as radiation, scattering, and material losses; PC: polarization controller; D1 & D2: photodetector. **b**, Sketch of how in **a** the CW mode couples to the CCW mode but the CCW mode does not couple to the CW mode, creating a nonreciprocal (unidirectional) coupling and thereby an exceptional surface (ES). Controlling the phase and the magnitude of the feedback allows steering the system on the ES. **c**, Without the feedback loop, the system has symmetric Lorentzian transmission (blue) and reflection (red) spectra for left and right incidence (CW and CCW inputs). Zero-reflection and the absence of mode splitting in the transmission spectra for both input directions imply the absence of intermodal coupling between the modes. The inset shows that the loading curves are also the same, and the system can be tuned from the overcoupling to the undercoupling regime through the critical coupling point.

Within the context of coupled-mode theory, our experimental setup is described by $\partial_t A = -iH_{ES}A$ where $A = (a_{cw}, a_{ccw})^T$, $a_{cw}$ and $a_{ccw}$ are the field amplitudes of the CW and CCW modes respectively, and $H_{ES}$ is the effective Hamiltonian given as

$$H_{ES} = \begin{pmatrix} \omega_0 - i\Gamma & 0 \\ \kappa & \omega_0 - i\Gamma \end{pmatrix}.$$

Here, $\omega_0 - i\Gamma$ are the complex frequencies of the degenerate CW and CCW modes, with $\Gamma = (\gamma_0 + \gamma_1)/2$ corresponding to the cavity loss rate which consists of the waveguide coupling loss $\gamma_1$ and all other resonator related losses (i.e., radiation, scattering and material absorption losses) $\gamma_0$, and $\kappa$ denotes the CW-to-CCW coupling strength. The zero value in the off-diagonal elements implies that the CCW mode does not couple to the CW mode. The unidirectional coupling strength in this system is defined as $\kappa = \alpha\gamma_1$ with $\alpha = |\alpha|\exp(i\phi)$ and $|\alpha|$ and $\phi$ corresponding to the magnitude of the feedback and phase acquired in the feedback loop. Both the eigenvalues and the corresponding eigenvectors of this system are degenerate and given as $\omega_{1,2} = \omega_0 - i\Gamma$ and $a_{1,2} = (0,1)^T$, forming an EP with CW chirality at the complex frequency $\omega_0 - i\Gamma$. Clearly, the system is at an EP for any non-zero $\kappa$ (i.e., $|\alpha| \neq 0$ and $\gamma_1 \neq 0$) and for all values of $\omega_0$ and $\Gamma$. Indeed, if $|\alpha|$ and $\phi$ are steered the system will trace an ES surface formed by EPs at the complex ES frequency $\omega_0 - i\Gamma$. As such, the system will always stay on a surface even if there are variations both in the amplitude and the phase of $\alpha$. Variations in $\gamma_0$ and $\gamma_1$, on the other hand, will create a new ES at a new complex ES frequency differing only in the imaginary part. Similarly, any perturbation (e.g., by temperature) that affects $\omega_0$ will lead to a new ES frequency differing only in the real part. Thus, although experimental imperfections and fluctuations may shift the complex ES frequency, they will not be able to lift the non-Hermitian degeneracy on the ES and the system will always remain on it.

We investigated the formation and the stability of an ES in our system by monitoring the reflection and transmission spectra for left and right incidence by tuning the system parameters $|\alpha|$ (feedback strength), $\phi$ (feedback phase), and $\gamma_1$ (waveguide-resonator coupling strength). We first set the system to critical coupling ($\gamma_0 = \gamma_1$), confirmed with zero transmission at the resonance dip both for left and right incidence. Then we vary $|\alpha|$ and $\phi$, implementing a variable/tunable reflector, and monitor transmission $T_{cw(ccw)}$ and reflection $R_{cw(ccw)}$. We observe symmetric transmission spectra $T_{cw} \equiv |t_{cw}|^2$ and $T_{ccw} \equiv |t_{ccw}|^2$ with Lorentzian lineshapes for all values of $|\alpha|$ and $\phi$. These results agree well with coupled mode theory which predicts $t_{cw(ccw)} \propto \Delta/(\Delta - i\gamma_0)$ and hence $T_{cw(ccw)} \propto 1/(1 + \gamma_0^2/\Delta^2)$, where $\Delta$ is the laser-cavity detuning. On the other hand, it is obvious that reflection spectra $R_{cw}$ and $R_{ccw}$ will demonstrate an asymmetric behaviour because $R_{ccw}$ is constant at all frequencies (i.e., for right incidence reflection occurs from the mirror only and does not involve the resonator) while $R_{cw}$ exhibits a resonance. This asymmetric reflection with symmetric transmission for inputs in opposite directions already indicates the presence of an EP. According to coupled mode theory, the amplitude reflection coefficient for light incident from the left scales as $r_{cw} \propto t_{cw}^2$, leading to $R_{cw} \equiv |r_{cw}|^2 \propto 1/(1 + \gamma_0^2/\Delta^2)^2$. In other words, the reflection spectrum for left incidence features a squared Lorentzian response. We observe this asymmetry and quartic behaviour (i.e., flattening) of the reflection lineshape in our experiments (see **Fig. 2a**). Following recent theoretical predictions[33], this type of response indicates the presence of a novel type of perfectly absorbing EP; lineshape modifications associated with such an EP have, however, not been observed up to date because, to the best of our knowledge, all previous experiments have been performed in the vicinity of an EP rather than exactly at an EP due to the difficulty of keeping a system at a discrete EP stably and continuously. Thus, the expected lineshape modification remained obscured and indiscernible in previous works. Our system, on the other hand, operates on an ES and is thus always exactly at an EP even in the presence of experimental imperfections and fluctuations.

To demonstrate that our system is, indeed, on the ES, we have collected $R_{cw}$ spectra at various values of $|\alpha|$ and $\phi$ when the system is at critical coupling ($\gamma_0 = \gamma_1$) and extracted the frequency and linewidth of the resonance lineshape on the ES. We do this by fitting the experimental data with a function composed of the product of two Lorentzians, that is $f = L_1 L_2$ with $L_k = A_k \Delta_k/(\Delta_k - i\Gamma_k)$, and by estimating $\{\Delta_k, \Gamma_k\}$ which should ideally satisfy $\Delta_1 = \Delta_2$ and $\Gamma_1 = \Gamma_2$ at an EP and on the ES. Plotting the experimentally obtained $\Delta w = \Delta_1 - \Delta_2$ and $\Delta\Gamma = \Gamma_1 - \Gamma_2$ as a function of $|\alpha|$ and $\phi$ has revealed the ES (see **Fig. 2b, c**). We have also performed experiments at undercoupling (**Fig. 3**) and overcoupling (**Fig. 4**) regimes by tuning $\gamma_1$ (i.e., varying the taper-resonator gap), and observed that the system always stays on an ES and remains robust against changes and unwanted fluctuations in the waveguide-resonator coupling strength. Thus, steering the system in the 2D parameter space using $|\alpha|$ and $\phi$ always defines an ES regardless of the waveguide-resonator coupling regime.

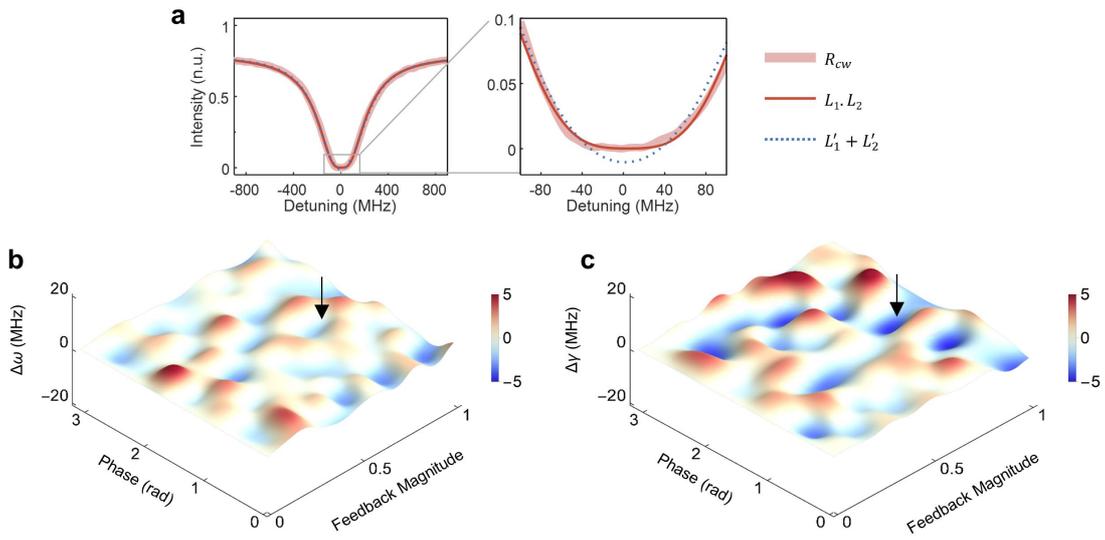

***Figure 2 | Squared Lorentzian reflection spectra and reconstructed exceptional surface for CW input at critical coupling. a,** Reflection spectrum $R_{cw}$ obtained for fully reflective end-mirror (feedback loop with $|\alpha| = 1$ and no BS coupler in Fig. 1a) has quartic lineshape (left panel). A closer look into the resonance dip of $R_{cw}$ reveals the flattening of the lineshape (right panel). Best curve fit is obtained using the function $f = L_1 \cdot L_2$ (product of two Lorentzians) rather than $f = L_1 + L_2$ (sum of two Lorentzians). **b,** Reconstructed ES in the 2D parameter space $\{|\alpha|, \phi\}$ of the system when the system is operated at the critical coupling for a CW input. Curve fitting to the experimentally obtained $R_{cw}$ is used to estimate the real and imaginary parts of the complex eigenfrequency of the system at the EP where two eigenfrequencies coalesce. $\Delta w$ (left panel) and $\Delta\gamma$ (right panel) correspond to the difference between the real and imaginary parts of two complex eigenfrequencies obtained from curve fitting. Deviations from zero are attributed to curve fitting noise. Arrows point to the locations of the spectrum on the ES.*

For all studied coupling regimes, The $\Delta w$ values are in the range $[-5.2 MHz, 5.3 MHz]$ and $\Delta\Gamma$ are in the range $[-5.6 MHz, 6 MHz]$, which, when normalized with the frequency and linewidth of the resonance at the critical coupling without the feedback, yield $|\Delta w/\omega_0| \lesssim 10^{-8}$ and $\Delta\Gamma/\Gamma_0 \lesssim 10^{-2}$, implying that the system is, indeed, on an ES. The system will leave the ES only for perturbations that break unidirectionality and establish a symmetric or asymmetric coupling between the CW and CCW modes. We also note that the system is on an ES at all resonances across the spectrum (i.e., one can construct and probe multiple ES in parallel by simultaneously probing multiple resonances).

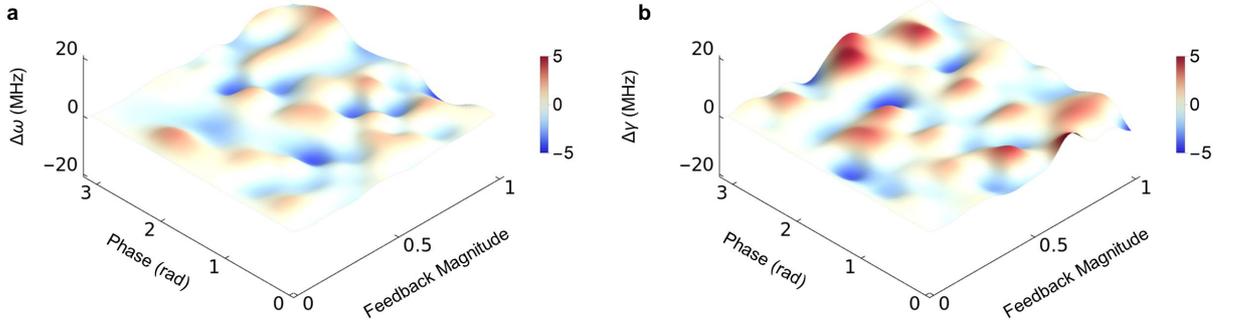

*Figure 3 | Experimentally obtained exceptional surface from the reflection spectra in the undercoupling regime. a, $\Delta w$ and b, $\Delta \gamma$ correspond to the difference between the real and imaginary parts of two complex eigenfrequencies obtained from curve fitting to $R_{cw}$ obtained at different phase $\phi$ and feedback strength magnitude $|\alpha|$.*

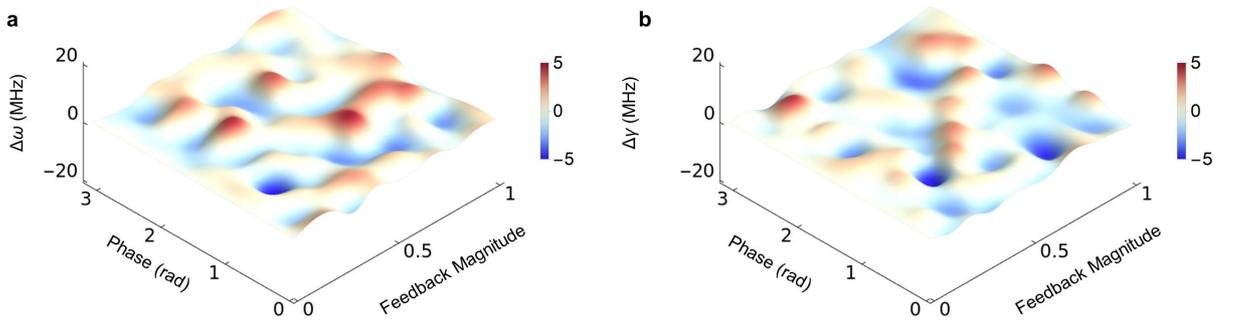

*Figure 4 | Experimentally obtained exceptional surface from the reflection spectra in the overcoupling regime. a, $\Delta w$ and b, $\Delta \gamma$ correspond to the difference between the real and imaginary parts of two complex eigenfrequencies obtained from curve fitting to $R_{cw}$ obtained at different phase $\phi$ and feedback strength magnitude $|\alpha|$.*

Next, we study the absorption properties of the system (see **Fig. 1a**) operating on the ES for left incidence (input in CW direction) at various taper-resonator coupling conditions when $|\alpha| = 1$ corresponding to a fully reflective system (see **Fig. 5**), where $T_{cw} = 0$. Under this condition, the absorption spectrum is calculated using $A_{cw} = 1 - R_{cw}$ where the reflection spectrum is measured by detector $D_1$. The normalization is carried out by considering the losses $L_{cw}$, including the insertion and propagation losses when the left incident field travels from the input circulator to the reflector and then back along the same path to $D_1$ in the absence of the resonator. The system stays on the ES as we vary the taper-resonator coupling strength, but the absorption strongly depends on the coupling regime, achieving perfect absorption only at the critical coupling (see **Fig. 5**). This absorption behaviour can be explained as follows: Our system operating at the critical coupling with $|\alpha| = 1$ (perfectly reflecting end-mirror) represents a one-channel coherent perfect absorber (CPA) with $T_{cw} = 0$ and $R_{cw} = 1/(1 + \gamma_0^2/\Delta^2)^2 \to 0$ for $\Delta \to 0$, and thus $A_{cw} = 1$ at the ES frequency, thus we refer to this special CPA as a CPA-ES. Different from a conventional one-channel CPA (or critical coupling), the resonator is tuned here to an EP on the ES and hence the absorption lineshape is quartic as is the reflection lineshape (see **Fig. 5**). As the system moves away from the critical coupling point towards the undercoupling or the overcoupling regime, a gradual transition from a squared Lorentzian lineshape to a more Lorentzian-like lineshape is clearly seen (**Fig. 5**). As discussed above our analysis compensates for the losses by taking them into account in the normalization process. If the off-resonance losses $L_{cw}$ are not accounted for, the absorption will be limited only by $L_{cw}$, that is $A_{cw} \to 1 - L_{cw}$ as $\Delta \to 0$, when the system is on the ES at the critical coupling.

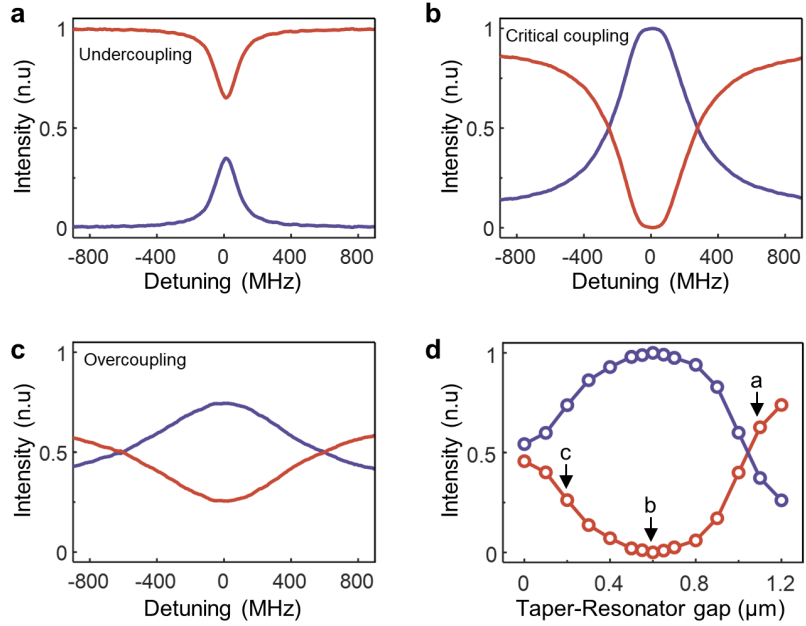

***Figure 5 | Chiral absorption and chiral coherent perfect absorption (CPA) on an exceptional surface with fully reflective end-mirror.*** *Absorption spectra $A_{cw}$ (blue curves) are inferred from the measured reflection spectra $R_{cw}$ (red curves) using $A_{cw} + R_{cw} = 1$ for a CW input and fully reflective feedback loop ($|\alpha| = 1$). Spectra measured at **a**, undercoupling regime, **b**, critical coupling, and **c**, overcoupling regime reveal a CPA-EP on the ES with a quartic absorption lineshape only for critical coupling, as in **b**. As the system moves away from the critical coupling regime the quartic behavior becomes indiscernible. **d**, Measured reflection and calculated absorption exactly at the ES frequency (i.e., zero detuning) for various taper-resonator coupling strength $\gamma_1$ determined by the gap between the taper and the resonator show the highest absorption at the critical coupling (i.e., gap equals to ~0.6μm) where we have $R_{cw} = 0$ and thus $A_{cw} = 1$. As the system moves from critical coupling towards undercoupling or overcoupling regimes, absorption at the resonance monotonously decreases. Circles are the values extracted from experimental data and the straight lines are inserted as guides to the eye. The labelled points in **d** are obtained from the spectra shown in **a, b,** and **c**.*

We now show that an ES in our system (see **Figs. 1a, 6**) leads to chiral absorption and CPA-ES also for non-ideal mirrors that are partially reflecting (see **Fig. 1a**). To demonstrate and probe these features, we performed experiments using a 10:90 end-mirror (10% transmission and 90% reflection) at different taper-waveguide coupling regimes. In these experiments, a portion of the transmission for CW input is monitored by $D_2$ with $R_{ccw}$ for right incidence being taken as the input power minus the transmitted power through the VOA. We have also measured losses $L_{cw} \neq L_{ccw}$ without the resonator involved and considered them in the normalization process (i.e., spectra are normalized with the power input to the tapered waveguide for left incidence and with the power just before the reflector for right incidence). We note that varying the feedback phase $\phi$ does not affect the observed features. Typical spectra obtained for the 10:90 end-mirror at different taper-waveguide coupling regimes are shown in **Fig. 6**.

When the end-mirror is not 100% reflecting, we have access to reflection and transmission spectra $(T_{cw}, R_{cw})$ and $(T_{ccw}, R_{ccw})$ for left (CW direction) and right incidence (CCW direction) from which the absorption spectra $A_{cw}$ and $A_{ccw}$ can be calculated using the expression $A_{cw(ccw)} + R_{cw(ccw)} + T_{cw(ccw)} = 1$. In the experiments, $T_{cw(ccw)}$ for left and right incidence exhibit typical resonance dips at the ES frequency with Lorentzian lineshapes. However, reflection spectra differed significantly: $R_{cw}$ has a squared Lorentzian lineshape with a flattened resonance dip around the ES frequency (see **Fig. 6**) whereas $R_{ccw}$ is constant ($e.g., R_{ccw} = 0.9$ for the 10:90 end-mirror) at all frequencies because it does not involve the resonator (see **Fig. 6**). This chiral behaviour (asymmetry in reflection) is the result of the strong coupling between the CW and CCW modes of the resonator only for left incidence (no coupling between them for right incidence), and it is the source of larger absorption for the left incidence compared to the right incidence, and hence of the chiral absorption. Indeed, because of this asymmetric reflection, the absorption spectrum $A_{cw}$ for CW input (left incidence) is a superposition of a Lorentzian term coming from the transmission $T_{cw}$ and a squared-Lorentzian term from $R_{cw}$ whereas $A_{ccw}$ for CCW input is always a Lorentzian. The weights of the Lorentzian and squared-Lorentzian terms in the superposition are determined by the reflectivity of the mirror (i.e., $A_{cw}$ is squared Lorentzian for a 100% reflecting mirror and it is Lorentzian for a 0% reflecting mirror) and hence the effective unidirectional coupling $\kappa$ between the CW and CCW modes. Another parameter that affects $\kappa$ and thereby the contribution of Lorentzian and squared-Lorentzian terms to the final lineshape is the waveguide-resonator coupling strength $\gamma_1$ through the expression $\kappa = \alpha \gamma_1$. Thus, when the taper-resonator coupling or the reflectivity of the end-mirror is varied, the system continues to be on an ES but the lineshape and overall amount of the absorption are significantly altered. Interestingly, a gradual transition from a quartic (squared-Lorentzian) form to a quadratic (Lorentzian) form in the $A_{cw}$ lineshape takes place as the taper-resonator coupling moves from the critical coupling towards the undercoupling or overcoupling regime (see **Fig. 6**, upper panel) or the reflectivity of the end-mirror is tuned. CPA-ES with flat-top squared Lorentzian lineshape is clearly seen when the system is at the critical coupling and the input is CW (see **Fig. 6**).

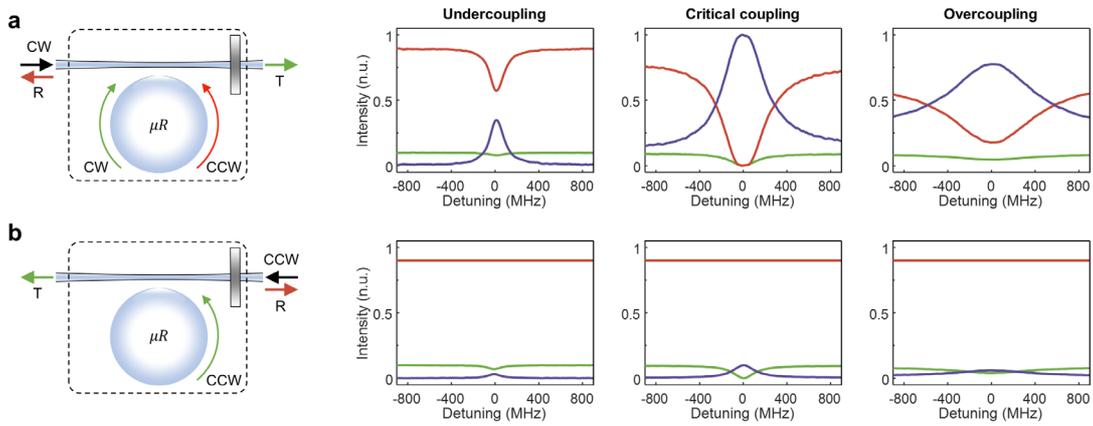

*Figure 6 | Chiral absorption and coherent perfect absorption (CPA) on an exceptional surface. Dotted boxes in the left panels in **a** and **b** represent the ES-device composed of a waveguide-coupled microresonator (μR) with an end-mirror with 10% reflection and 90% transmission. Black arrows denote the CW and CCW input ports of the ES-device, and red and green arrows represent the corresponding reflection and transmission ports. In the case of CW input as in **a**, the field inside μR has both CW and CCW components whereas it has only CCW component for the CCW input as in **b**. Measured transmission $T_{cw(ccw)}$ (green) and reflection $R_{cw(ccw)}$ (red) spectra and calculated absorption $A_{cw(ccw)} = 1 - R_{cw(ccw)} - T_{cw(ccw)}$ (blue) spectra of the ES-device at the undercoupling, critical coupling and overcoupling regimes for CW (upper panel) and CCW (lower panel) inputs. $T_{cw}$ and $T_{ccw}$ have Lorentzian lineshapes with resonance dips at zero-detuning (ES frequency) at all coupling regimes; $R_{ccw}$ is constant at all frequencies; and $R_{cw}$ exhibits squared Lorentzian spectra. CPA-ES with quartic lineshape is observed at the critical coupling for CW input only implying chiral CPA. $A_{cw}$ is at least ten times larger than $A_{(ccw)}$, and hence chiral absorption at all coupling conditions.*

Finally, we steer the system on the ES and determine the absorption $A_{cw}$ and $A_{ccw}$ at the ES frequency for the left (see **Fig.7a**) and the right (see **Fig.7b**) incidence, respectively, by tuning the VOA drive voltage $V_{drive}$ and the taper-resonator coupling. In this experiment, the BS before the feedback loop (see **Fig. 1c**) is a 10:90 beamsplitter and the VOA in the feedback loop controls the back-reflection into the CCW mode for left incidence (CW input). Thus, the reflectivity $|\alpha|$ is varied between 0 and 0.9, i.e. $|\alpha| = 0.9$ at $V_{drive} = 0V$ and $|\alpha| = 0$ at $V_{drive} = 1V$. For right incidence (CCW input), the 10:90 BS does not play a role and the VOA simulates an end-mirror with reflectivity $0 \leq |\alpha| \leq 1$, i.e. $|\alpha| = 0$ at $V_{drive} = 1V$ corresponding to $R_{ccw} = 0$ and $|\alpha| = 1$ at $V_{drive} = 0V$. We present the results for $0 \leq |\alpha| \leq 0.9$ in **Fig. 7** demonstrating that CPA-ES ($A_{cw} = 1$ at resonance with squared-Lorentzian lineshape) takes place for left incidence at the critical coupling for all values of $|\alpha|$ (**Fig.7a**). For right incidence, on the other hand, conventional CPA ($A_{ccw} = 1$ at resonance with Lorentzian lineshape) occurs at the critical coupling only for $|\alpha| = 0$, and $A_{ccw}$ decreases with increasing $|\alpha|$. As the taper-resonator gap increases from zero (i.e., overcoupling), $A_{cw}$ and $A_{ccw}$ first increase reaching its maximum value at the critical coupling, and then starts decreasing as the gap increases (the system moves towards deep undercoupling regime). Chirality of absorption can be better seen in the ratio $\xi = A_{ccw}/A_{cw}$ (**Fig. 7c**) which can be tuned in the range $[0,1]$: $\xi = 0$ denotes no absorption for right incidence ($A_{ccw} = 0$) corresponding to a fully reflecting end-mirror; $\xi = 1$ denotes equal absorption for incidence in both directions ($A_{ccw} = A_{cw}$); and all other values of $0 < \xi < 1$ imply chiral absorption, that is $A_{cw} > A_{ccw}$, with higher chirality at higher $|\alpha|$ and at taper-resonator gap closer to the critical coupling condition.

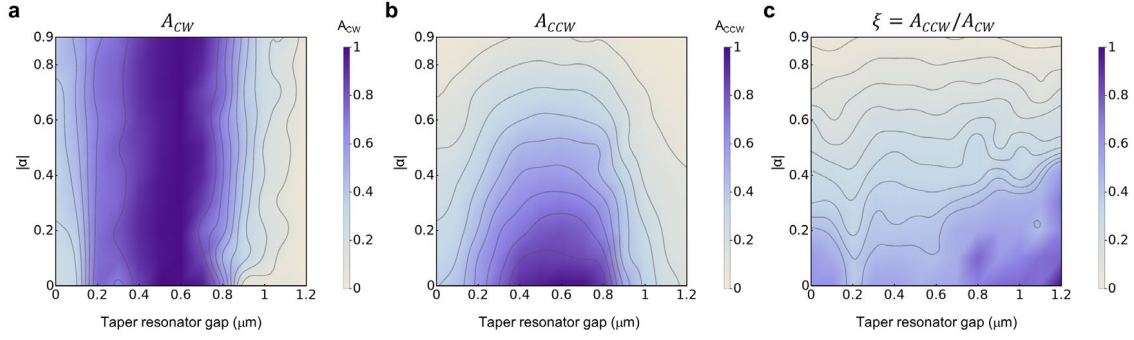

*Figure 7 | Tunable chiral absorption on an exceptional surface by varying reflection from the end-mirror or the waveguide-resonator coupling strength. a, b, The amount of absorption at the ES frequency for left (CW input) and right (CCW input) incidence can be tuned by controlling taper-resonator gap and the reflectivity $|\alpha|$ of the end-mirror, implemented with the VOA in the feedback loop. c, Absorption ratio $\xi = A_{ccw}/A_{cw}$ of right incidence to left incidence at the ES frequency. CPA-ES is obtained at the critical coupling for left incidence. Critical coupling is achieved when the taper-resonator gap is 0.6 µm, with the gap smaller than 0.6 µm corresponds to overcoupling and a gap larger than 0.6 µm corresponds to undercoupling.*

It is worth noting that ES does not only allow chiral absorption at the ES frequency but also provides a way to control the absorption bandwidth. We have observed that the absorption bandwidth defined as full-width at half-maximum is different for left and right incidence (i.e., chirality in bandwidth). More interestingly, we have found that compared to the conventional CPA at critical coupling, the absorption bandwidth at the CPA-ES is 1.18, 1.53, and 1.58 times larger for 50:50, 10:90, and fully reflecting end-mirrors, respectively. These values are close to the theoretically predicted values of 1.27, 1.50, and 1.55. The flattened absorption spectrum in the vicinity of critical coupling for left incidence may provide a remedy to the narrow absorption bandwidth of a conventional CPA—a problem that has hindered progress in technologies relying on CPA.

In conclusion, we demonstrate a non-Hermitian optical device which exhibits exceptional surface (ES) and chiral absorption (CPA). Since the device operates always exactly at an EP when on the ES, it provides a stable and controllable platform to study EP-related phenomena and processes, revealing previously unobserved features such as a CPA-ES with quartic lineshape and chiral absorption on the ES. Our results will pave the way towards control of various optical processes and light-matter interactions exactly at an EP (not limited to the spectra in the vicinity of an EP), with potential applications ranging from chiral light-matter interaction, lasing, and emission to chiral nonlinear photonics and photovoltaics. Creating ES through a simple unidirectional coupling route between two modes can be extended to other physical platforms where the coupling between modes and systems can be made unidirectional by electrical, optical, photonics or acoustic feedback. Since no additional loss or gain is introduced into the system, ES obtained through unidirectional coupling can also benefit studies of quantum dynamics in non-Hermitian systems.

**References**

[1] T. Kato, Perturbation Theory of Linear Operators (Springer, Berlin, 1966).

[2] Özdemir, Ş. K., Rotter, S., Nori, F. & Yang, L. Parity-time symmetry and exceptional points in photonics. *Nat. Mater.* **18,** 783-798 (2019).

[3] El-Ganainy, R. *et al.* Non-Hermitian physics and PT symmetry. *Nat. Phys.* **14**, 11–19 (2018).

[4] Miri, M. A. & Alù, A. Exceptional points in optics and photonics. *Science* **363**, eaar7709 (2019).

[5] Feng, L., El-Ganainy, R. & Ge, L. Non-Hermitian photonics based on parity-time. *Nat. Photon*. **11**, 752–762 (2017).

[6] El-Ganainy, R., Khajavikhan, M., Christodoulides, D. N. & Özdemir, Ş. K. The dawn of non-Hermitian optics. *Commun. Phys.* **2**, 37 (2019).

[7] Berry, M. V. M. The Adiabatic Phase and Pancharatnam's Phase for Polarized Light. *J. Mod. Opt*. **34**, 1401-1407 (1987).

[8] Yarkony, D. R. Diabolical conical intersections. *Rev. Mod. Phys.* **68**, 985–1013 (1996).

[9] Peng, B. et al. Chiral modes and directional lasing at exceptional points. *Proc. Natl. Acad. Sci. U.S.A.* **113,** 6845-6850 (2016).

[10] Chen, W., Özdemir, Ş. K., Zhao, G., Wiersig, J. & Yang, L. Exceptional points enhance sensing in an optical microcavity. *Nature* **548**, 192-196 (2017).

[11] Hodaei, H. *et al.* Enhanced sensitivity at higher-order exceptional points. *Nature* **548**, 187–191 (2017).

[12] Lau, H.-K. & Clerk, A. A. Fundamental limits and non-reciprocal approaches in non-Hermitian quantum sensing. *Nat. Commun.* **9**, 1–13 (2018).

[13] Lai, Y.-H., Lu, Y.-K., Suh, M.-G., Yuan, Z. & Vahala, K. Observation of the exceptional-point-enhanced Sagnac effect. *Nature* **576**, 65–69 (2019).

[14] Hokmabadi, M. P., Schumer, A., Christodoulides, D. N. & Khajavikhan, M. Non-Hermitian ring laser gyroscopes with enhanced Sagnac sensitivity. *Nature* **576**, 70–74 (2019).

[15]     Park, JH., Ndao, A., Cai, W. *et al*. Symmetry-breaking-induced plasmonic exceptional points and nanoscale sensing. *Nat. Phys.* **16**, 462–468 (2020).

[16]     Guo, A. *et al*. Observation of PT-Symmetry Breaking in Complex Optical Potentials. *Phys. Rev. Lett.* **103**, 093902 (2009).

[17]     Peng, B. *et al.* Loss-induced suppression and revival of lasing. *Science* **346**, 328–332 (2014).

[18]     Brandstetter, M. *et al*. Reversing the pump dependence of a laser at an exceptional point. *Nat. Commun*. **5**, 4034 (2014).

[19]     Bender, C. M., and Boettcher, S. Real spectra in non-Hermitian Hamiltonians having PT symmetry. *Phys. Rev. Lett.* **80**, 5243–5246 (1998).

[20]     Peng, B. *et al*. Parity-time-symmetric whispering-gallery microcavities. *Nat. Phys.* **10**, 394-398 (2014).

[21]     Rüter, C. E. *et al*. Observation of parity-time symmetry in optics. *Nat. Phys.* **6**, 192-195 (2010).

[22]     Jing, H., Ozdemir, S. K., Lu, X.-Y., Zhang, J., Yang, L. and Nori, F. PT-symmetric phonon laser. *Phys. Rev. Lett.* **13**, 053604 (2014).

[23]     Zhen, B. *et al*. Spawning rings of exceptional points out of Dirac cones. *Nature* **525**, 354-358 (2015).

[24]     Zhang, J. *et al*. A phonon laser operating at an exceptional point. *Nat. Photon.* **12**, 479–484 (2018).

[25]     Naghiloo, M., Abbasi, M., Joglekar, Y.N. & Murch, K.W. Quantum state tomography across the exceptional point in a single dissipative qubit. *Nat. Phys.* **15**, 1232-1236 (2019).

[26]     Xiao, L. *et al*. Observation of critical phenomena in parity-time-symmetric quantum dynamics. *Phys. Rev. Lett.* **123**, 230401 (2019).

[27] Xu, H., Jiang, L., Clerk, A.A. & Harris, J.G.E. Nonreciprocal control and cooling of phonon modes in an optomechanical system. *Nature* **568**, 65-69 (2019).

[28] Zhong, Q. et al. Sensing with exceptional surfaces in order to combine sensitivity with robustness. *Phys. Rev. Lett.* **122,** 153902 (2019)

[29] McDonald, A. & Clerk, A.A. Exponentially-enhanced quantum sensing with non-Hermitian lattice dynamics. *Nat Commun* **11**, 5382 (2020).

[30] Zhong, Q., Hashemi, A. , Ozdemir, S. K., and El-Ganainy, R. Control of spontaneous emission dynamics in microcavities with chiral exceptional surfaces. *Phys. Rev. Research* **3**, 013220 (2021).

[31] M. Khanbekyan and J. Wiersig, Decay suppression of spontaneous emission of a single emitter in a high-Q cavity at exceptional points. *Phys. Rev. Research* **2**, 023375 (2020).

[32] Q. Zhong, S.K. Ozdemir, A. Eisfeld, A. Metelmann, and R. El-Ganainy. Exceptional-Point-Based Optical Amplifiers. *Phys. Rev. Applied* **13**, 014070 (2020).

[33] Sweeney, W. R., Hsu, C. W., Rotter, S. & Stone, A. D. Perfectly Absorbing Exceptional Points and Chiral Absorbers. *Phys. Rev. Lett.* **122**, 093901 (2019).

[34] Zhong, Q., Nelson, S., Özdemir, Ş. K. & El-Ganainy, R. Controlling directional absorption with chiral exceptional surfaces. *Opt. Lett.* **44**, 5242-5245 (2019).

**Acknowledgements**

This research is supported by Air Force Office of Scientific Research (AFOSR) Multidisciplinary University Research Initiative (MURI) Award No. FA9550-21-1-0202. S.K.O. also acknowledges support from NSF (Grant No. ECCS 1807485) and AFOSR (Award no. FA9550-18-1-0235). R.E. acknowledges support from ARO (Grant No.W911NF-17-1-0481), NSF (Grant No. ECCS 1807552), and the Alexander von Humboldt Foundation. S.R. acknowledges funding by the European Commission (MSCA-RISE 691209) and by the Austrian Science Fund (FWF) (P32300).

**Contributions**

S.K.O., R.E., and S.R. conceived the idea and supervised the research. S.K.O., S.S. and R.E. designed the experiments; S.S. performed the experiments with help from M.M. at the initial stages of the study; S.S. collected the experimental data and analyzed it together with Q.Z.; S.S. and Q.Z. performed numerical simulations with guidance from R.E., S.R., and S.K.O.; S.K.O. and R.E. wrote the manuscript with contributions from all authors. All authors read and agreed with the content and discussions in the manuscript.

**Corresponding authors**

Correspondence to Sahin K. Ozdemir (E-mail: sko9@psu.edu).